\UseRawInputEncoding
\documentclass[pre,aps,showpacs,10pt,reprint,superscriptaddress,longbibliography]{revtex4-1}
\usepackage[mathletters]{ucs}
\usepackage[utf8x]{inputenc}
\usepackage{graphicx}
\usepackage{amsmath}
\usepackage{amsfonts}
\usepackage{amssymb,yfonts,bbm}
\usepackage[bbgreekl]{mathbbol}
\usepackage{mathrsfs}   
\usepackage{color}					
\usepackage{bm}      				
\usepackage{chemarrow}				
\usepackage{booktabs}				
\usepackage{array}					
\usepackage[bbgreekl]{mathbbol}  	
\usepackage{slashed}  				
\usepackage[hidelinks]{hyperref}	
\hypersetup{colorlinks=true,citecolor=blue,urlcolor=blue} 
\usepackage{tikz}					
\usepackage{verbatim}
\usepackage{notes2bib}				
\bibnotesetup{note-name =,use-sort-key = false}

\DeclareMathAlphabet{\mathpzc}{OT1}{pzc}{m}{it}
\usepackage{svg}        

\newcolumntype{C}[1]{>{\centering\arraybackslash$}m{#1}<{$}}

\begin{document}

\title{Thermodynamic Circuits I: Association of  devices in stationary nonequilibrium}

\author{Paul Raux}
\affiliation{Université Paris-Saclay, CNRS/IN2P3, IJCLab, 91405 Orsay, France }
\affiliation{Université Paris Cité, CNRS, LIED, F-75013 Paris, France}
\author{Christophe Goupil}
\affiliation{Université Paris Cité, CNRS, LIED, F-75013 Paris, France}
\author{Gatien Verley}
\affiliation{Université Paris-Saclay, CNRS/IN2P3, IJCLab, 91405 Orsay, France }

\date{\today}

\begin{abstract}
For a circuit made of thermodynamic devices in stationary nonequilibrium, we determine the mean currents (of energy, matter, charge, etc) exchanged with external reservoirs driving the circuit out of equilibrium. Starting from the conductance matrix describing the nonlinear current--force characteristics of each device, we obtain the conductance matrix of the composite device. This generalizes the rule of resistance addition (serial association) or conductance addition (parallel association) in stationary out-of-equilibrium thermodynamics and for multiple coupled potentials and currents of different natures. Our work emphasizes the pivotal role of conservation laws when creating circuits of complex devices. Finally, two examples illustrate the determination of the conservation laws for the serial and parallel associations of thermodynamic devices.
\end{abstract}

\maketitle

\section{Introduction} 

Dividing a problem into several pieces often simplifies its resolution. Combining this approach with a graphical representation produces circuits with components of lower complexity. The circuit's connections conveniently summarize the conservation of physical quantities (energy, momentum, charge, chemical species, etc.) circulating through the components. Different kinds of circuits or graphs exist: bond graphs in engineering science~\cite{Thoma2016vol}, Feynman diagrams in particle physics~\cite{Schwartz2013vol}, electric circuits in electrokinetic~\cite{Horowitz2015vol, Bringuier2005vol} or hyper-graphs of chemical reactions in chemistry~\cite{Feinberg2019vol, DalCengio2023vol13}. Electric power and signal processing represent the paramount application of circuit theory, although only one conserved quantity is usually considered. In this case of a single (or multiple but decoupled) potential(s), the problem has long since been solved. The case of coupled potentials and general boundary conditions has drawn many works~\cite{ Kedem1965_vol61, Katchalsky1965vol, Peusner1986vol5, Peusner1985vol83, Thoma1977vol303}. Neumann (fixed current) or Dirichlet (fixed potential) conditions are often assumed but less often mixed boundary conditions where both currents and potentials achieve non-prescribed stationary values. For those, real integrated balances at the boundaries are central in determining the stationary state. A nonlinear theory dealing with mixed boundary conditions and several conserved quantities coupled through complex circuits is appealing. A step in this direction was recently taken in Ref.~\cite{Avanzini2023vol13} for chemical reaction networks: elementary reactions are gathered in submodules with an effective description involving only the species exchanged with the reservoirs or other modules. Then, the current--concentration characteristics for those species are sufficient to determine the working point of the circuit associating all modules and the stationary currents exchanged with its environment. However, eliminating all internal degrees of freedom and going beyond multivariable functions connecting currents and forces within global characteristics call for more operational and systematic methods valid for any thermodynamic system. This series of articles on Thermodynamic Circuits describes some achievements in this direction.

The foundation of a circuit theory for thermodynamic devices is now possible by relying on first, the convenient use of conservation laws to reduce the number of physical currents to a fundamental set~\cite{Polettini2014_vol141,  Polettini2016_vol94}; second on the introduction of nonequilibrium conductance matrices to characterize thermodynamic devices and their current--force relations accurately. The conductance matrices are the multidimensional generalization of the scalar current--force characteristic of dipoles (e.g., current--voltage for an electric dipole). As such, they extend to vectors the concept of resistance defined as the scalar ratio between a force and a current. 
Conductance matrices generalize to nonequilibrium stationary states Onsager's response matrices of linear irreversible thermodynamics~\cite{Onsager1931_vol37, Onsager1931_vol38, Callen1985_vol}. It is possible to compute them when microscopic modeling is available, as recalled in section \ref{NEcond} of Thermodynamic Circuits~I. It is also possible to obtain them by integrating a local response, although some freedom remains beyond the linear case, as we argue in the second article of this series. In any case, choosing nonequilibrium conductance matrix is a matter of device modeling that includes the description of currents up to second cumulants~\cite{Vroylandt2018vol2018}. Such a modelization will become fully operational when a method of direct measurement of nonequilibrium conductances is available. 
In section \ref{Serial Association} and \ref{Parallel Association} of Thermodynamic Circuits~I, we present our general theory for the association of thermodynamic devices in stationary nonequilibrium. We aim to provide the nonequilibrium conductance matrix of a composite thermodynamic system using its two subsystems' and their conservation laws. 
The serial association is studied in section \ref{Serial Association}. First, we state the problem of mixed boundary conditions. We explain how current conservation at the interface allows, in principle, the integration of the internal degrees of freedom, i.e., the local potentials at the connection. Second, we generalize the law of resistance addition to a law of resistance matrix addition. Conservation laws within each subsystem are used to match the matrix dimension of the conductances to the one of the composite device. 
The parallel association is studied in section \ref{Parallel Association}. We generalize the law of conductance addition since the connected devices remain in Dirichlet boundary conditions: There is no need to determine internal degrees of freedom here. First, we obtain the conductance matrix at the level of physical currents (linearly dependent through conservation laws) and forces. Second, we determine the conservation laws for the parallel association of the two sub-devices. From those, the conductance matrix at the fundamental currents level can be computed on any desired basis. We illustrate our results on two examples chosen to apply to the most general cases of serial and parallel associations.

The other articles of this series present various applications of our main results.  The paradigmatic case of thermoelectric converters (TEC) is considered in Thermodynamic Circuits~II. There, we develop our nonlinear framework for thermo-electricity using the concept of nonequilibrium conductance and study the various optima of a TEC. In Thermodynamic Circuits~III, we consider the serial and parallel association of two different TECs and observe the consequences of the mixed boundary conditions on the thermo-electric conversion in the serial case. In Thermodynamic Circuits~IV, we apply our framework to the association of chemical reaction networks described using deterministic dynamics. We introduce the nonequilibrium conductance matrix of the subnetworks and determine the one of the composite network.

\section{Nonequilibrium conductance matrix for Markov processes}
\label{NEcond}

This section gives a self-contained derivation of the nonequilibrium conductance matrix for an autonomous device modeled by a Markov jump process driven out of equilibrium by competing external reservoirs. This summarizes some results of Ref~.\cite{Vroylandt2018vol2018}. The concept of the nonequilibrium conductance matrix is not restricted to but is most precisely defined within this framework. 

Let us consider a system with a finite set $\mathscr{V}$ of states  generically denoted by $x,y,\dots$ in $\mathscr{V}$. The cardinal of sets is generically denoted with $|\dots|$; hence, $|\mathscr{V}|$ is the number of states. The system's dynamics is a Markov jump process on $\mathscr{V}$. The transition rate from state $y$ to state $x$ via a channel $\nu$ is denoted $k_{(xy,\nu)}$. Then, $ k_{(xy,\nu)}dt$ is the probability of this transition to occur during $dt$ given that the system is initially in state $y$. Graphically, the states are represented as vertices and transitions as edges of a connected graph. Several channels mean that two vertices can be connected by more than one edge. We denote by $\mathscr{E}$ the set of all edges and by $e\in \mathscr{E}$ a specific edge $e=(xy,\nu)$. The stationary probability vector $\bm p^\mathrm{st}$ is the right null eigenvector of the Markov matrix $\bm k$ with off-diagonal components $(\bm k )_{xy} = \sum_\nu k_{(xy,\nu)}$. 

\subsection{Decomposition of stationary currents and forces}
\label{descriptionlevel}

In this section, we write the Entropy Production Rate (EPR) $\sigma$ as the sum, over all conjugated pairs, of the current-force products. This can be done at several levels of description from microscopic to macroscopic variables, providing on the way the relations enabling to switch from one level of description to another~\cite{Schnakenberg1976_vol48,Polettini2014_vol141,Polettini2016_vol94}.  

\subsubsection{Microscopic level} 
In the stationary state, the edge force $f_e$ along edge $e=(xy,\nu)$ is the logarithm of the ratio of forward--backward transition probabilities
\begin{equation}
	f_{(xy,\nu)} \equiv \ln \frac{k_{(xy,\nu)}p^\mathrm{st}_{y}}{k_{(yx,\nu)}p^\mathrm{st}_{x}}.
\end{equation}
The edge probability current along edge $e=(xy,\nu)$ is the difference of forward--backward transition probabilities
\begin{equation}
	j_{(xy,\nu)} \equiv k_{(xy,\nu)} p^\mathrm{st}_y - k_{(yx,\nu)} p^\mathrm{st}_x.
\end{equation}
Given the conservation of probability currents at each vertex of the transition graph, it is unsurprising that those currents are linearly dependent (Kirchhoff currents law at each vertex). Let us define the incidence matrix $\bm D$, with $|\mathscr{V}|$ rows and $|\mathscr{E}|$ columns, by its components
\begin{equation}
	D_{x,e} = \left \{ \begin{array}{l} +1 \text{ if } x \text{ is the target state of edge } e, \\ -1 \text{ if } x \text{ is the source state of edge } e, \\ 0 \text{ otherwise. } \end{array} \right. \label{IncidenceMatrix}
\end{equation}
By construction, the conservation of probability currents writes $\bm D \bm j = 0$ where $\bm j$ is the vector of components $j_{e}$. The matrix $\bm {\mathcal{C}}$ whose columns are a basis of the kernel of $\bm D$ is called the matrix of fundamental cycles~\cite{Schnakenberg1976_vol48}. Its $|\mathscr{C}|$ columns provide a basis for the vector space of cycles, where $\mathscr{C}$ is a set of integers labeling fundamental cycles. Among all edge probability currents, one can select linearly independent currents $\bm J$, called cycle currents, such that 
\begin{equation}
\bm j = \bm{\mathcal{C}} \bm J.
\end{equation}
This uses the Kirchoff currents law to deal with a reduced set of currents only. We remark that $\bm D \bm j = \bm D \bm {\mathcal{C}} \bm J  = 0$ is satisfied since $\bm D \bm {\mathcal{C}} = 0$. Furthermore, we define the cycle force by the sum of the forces on the edges belonging to the considered cycle. Then, the vector of cycle forces reads
\begin{equation}
	\bm F^{T} = \bm f^\mathrm{T} \bm{\mathcal{C}}
\end{equation}
where superscript $\mathrm{T}$ indicates transposition and $\bm f$ is the vector of components $f_{e}$. This choice of cycle forces guarantees the unicity of the entropy production rate expressed using edge or cycle variables
\begin{equation}
	\sigma = \bm f^\mathrm{T} \bm j = \bm f^\mathrm{T}  \bm{\mathcal{C}} \bm J = \bm F^\mathrm{T} \bm J.
\end{equation}
Let us emphasize that the choice of fundamental cycle is not unique, leading to many ways of expressing the EPR. In any case, the use of probability currents conservation is central for focusing on independent currents and is associated with the Euler formula of graph theory \cite{Schnakenberg1976_vol48}
\begin{equation}
	|\mathscr{E}|=|\mathscr{V}|-1 + |\mathscr{C}|.
\end{equation}
Indeed, Kirchhoff currents law at $|\mathscr{V}|-1$ vertices is enough to guarantee probability currents conservation at all vertices of the graph. Then, the number of independent currents among the $|\mathscr{E}|$ edge currents is $|\mathscr{C}|$ as expected.

\subsubsection{Macroscopic level} 
\label{Macro}
The above reasoning applies equally at the macroscopic level. Indeed, the physical currents exchanged with each reservoir are linearly dependent due to various conservation laws, just as probability currents are conserved at each vertex via Kirchhoff currents law. Let us first define the physical currents as 
\begin{equation}
	i_{p} = \sum_{c \in \mathscr{C}} \phi_{p c}J_{c}  \quad \Rightarrow \quad 
	\bm i = \bm{\phi} \bm J.
\end{equation}
where $p \in \mathscr{P}$ the set of integers labeling physical currents.
We denote by $J_{c}$ the $c$th component of the cycle current vector $\bm J$ and $\phi_{p c}$ the amount of $p$th physical quantity received by the system when performing cycle $c$. The local potentials vector $\bm a$ of component $a_{p}$ associated with the $p$th reservoir are related to cycle forces by
\begin{equation}
	\bm F^\mathrm{T} =  \bm a^\mathrm{T} \bm{\phi}, 
\end{equation}
so that the \emph{EPR} remains the same whether expressed using edges, cycles, or physical currents and their conjugated forces:
\begin{equation}
	\sigma = \bm F^\mathrm{T} \bm J = \bm a^\mathrm{T}  \bm{\phi} \bm J = \bm a^\mathrm{T}  \bm i .
\end{equation}

Due to the conservation of physical quantities (such as energy and matter),  physical currents $ i_{p}$ are linearly dependent. We denote the set of conservation laws $\mathscr{L} =\{\bm \ell_{1},\bm \ell_{2},\dots  \}$. Each element in this set is a row vector $\bm \ell_k$ with components equal to 0 or 1. The conservation law vectors $\bm \ell_k$ are linearly independent and satisfy by definition $\forall k, \;\bm \ell_{k}\bm i = \bm 0$. We denote $\bm \ell$ the matrix whose $k$th line is $\bm{\ell}_{k}$. For later convenience, we number the pins with consecutive positive integers when their associated currents belong to the same conservation law. Now, we can choose a maximal subset $\mathscr{I} \subset \mathscr{P}$ of physical currents that are linearly independent. These currents appear in the so-called vector of fundamental currents denoted $\bm I$ from which any physical currents can be recovered from
\begin{equation}
	  \bm i = \bm S \bm I, \label{iSI}
\end{equation}
where the selection matrix $\bm S$ has $|\mathscr{I}|$ independent columns belonging to the kernel of $\bm \ell$, i.e. $\bm \ell \bm S = 0$. From the columns independence, a unique Moore-Penrose inverse $\bm S^{+} \equiv [\bm S^{T} \bm S]^{-1} \bm S^{T}$ exists and leads to
\begin{equation}
	  \bm S^{+} \bm i =  \bm I. \label{S+iI}
\end{equation}
Hence, choosing a selection matrix $\bm S$ is equivalent to choosing fundamental currents. The fundamental forces conjugated to fundamental currents are defined by
\begin{equation}
	\bm A^\mathrm{T} = \bm a^\mathrm{T} \bm S, \label{FundForce}
\end{equation}
where again the \emph{EPR} is the same at the physical and fundamental levels
\begin{equation}
	\sigma = \bm a^\mathrm{T}  \bm i  = \bm a^\mathrm{T} \bm S \bm I =  \bm A^\mathrm{T} \bm I  .
\end{equation}
The choice of fundamental currents and forces is not unique in general, leading to many ways of expressing the EPR. In any case, conservation laws are central to focus on a set of fundamental currents. The number of physical currents verifies
\begin{equation}
	|\mathscr{P}|=|\mathscr{L}| + |\mathscr{I}|.
\end{equation}
which is the macroscopic equivalent of the Euler formula at the microscopic level. In table~\ref{notation-current-force}, we summarize the various notations for the currents and forces used in this section. 
\begin{table} \centering
	\begin{tabular}{|c||l|l|}  \hline
					& Full set & Reduced set \\ \hline \hline
		Micro 	& Edge current $\bm j$ & Cycle current $\bm J$ \\
		 & Edge force $\bm f$ &  Cycle force $\bm F$  \\
		 & Vector dimension  $|\mathscr{E}|$ & Vector dimension  $|\mathscr{C}|$  \\ \hline 
		Macro 	& Physical current $\bm i$ & Fundamental current $\bm I$ \\
		 & Local potential $\bm a$ & Fundamental force $\bm A$ \\
		 & Vector dimension  $|\mathscr{P}|$ & Vector dimension  $|\mathscr{I}|$ \\ \hline
	\end{tabular}
	\caption[Currents and forces at different levels of description.]{Notation for the currents and forces at different levels of description, from the microscopic level of edges and cycles to the macroscopic level of physical currents exchanged with the environment and the reduced set of fundamental currents. Lower case letters are for the full set of variables and upper case letters are for the reduced set of variables. We follow here the usual convention that $\bm j$ are local currents and $\bm i$ are integrated currents. One switches from the full set to the reduced set of variables using the $|\mathscr{V}|-1$ Kirchoff's current law (Micro) or the $|\mathscr{L}|$ conservation laws (Macro). \label{notation-current-force}}
\end{table}

\subsection{Nonlinear conductance matrices}

The relations between currents (or between forces) obtained in the previous section lead to a nonlinear conductance (or resistance) matrix at any level of description. We start at the microscopic level by defining a trivial relation between edge forces and currents through an edge resistance matrix. This matrix is chosen diagonal, hence all current correlations are neglected at this level. Then, we propagate this resistance matrix to higher levels of description up to the macroscopic level of fundamental currents and forces. 

First, we define the edge resistance along $e$ by
\begin{equation}
	r_e \equiv \frac{f_e}{j_e} > 0,
 \label{eq : resistance de lien}
\end{equation}
so that the edge force is given by $\bm f = \bm r \bm j$. The cycle resistance matrix defined by
\begin{equation}
	\bm{R} = \bm{\mathcal{C}}^\mathrm{T}\bm r \bm{\mathcal{C}}
 \label{eq : resitance cycles}
\end{equation}
relates cycle currents and forces since
\begin{equation}
	\bm F = \bm{\mathcal{C}}^\mathrm{T}\bm f =  \bm{\mathcal{C}}^\mathrm{T}\bm r \bm j =  (\bm{\mathcal{C}}^\mathrm{T}\bm r \bm{\mathcal{C}}) \bm J = \bm{R} \bm J.
\end{equation}
From Eqs.~(\ref{eq : resistance de lien}--\ref{eq : resitance cycles}), the matrix $\bm{R}$ is a square symmetric matrix that is positive definite in agreement with the positivity of the \emph{EPR}. Hence, there is an inverse matrix called cycle conductance matrix $ \bm{R}^{-1}$ such that
\begin{equation}
	\bm J = \bm{R}^{-1} \bm F.
\end{equation}
Proceeding similarly at the level of physical currents and forces leads to the physical conductance matrix 
\begin{equation}
	\bm g = \bm \phi \bm{R}^{-1} \bm\phi^\mathrm{T}
\end{equation}
since
\begin{equation}
	\bm i  = \bm \phi \bm J = \bm \phi \bm{R}^{-1} \bm F = (\bm \phi \bm{R}^{-1} \bm\phi^\mathrm{T}) \bm a = \bm g \bm a,
\end{equation}
and to the fundamental conductance matrix 
\begin{equation}
	\bm G = \bm S^{+} \bm g \bm S^{+\mathrm{T}},
\end{equation}
since
\begin{equation}
	\bm I  = \bm S^{+} \bm i =  \bm S^{+} \bm g \bm a = (\bm S^{+} \bm g \bm S^{+\mathrm{T}}) \bm A = \bm G \bm A.
\end{equation}
We end this section stressing that the nonequilibrium conductance matrix is a function of the macroscopic forces $\bm G = \bm G(\bm A)$, explaining the non-linearity of the current-force relation. Although introduced above in the framework of Markov jump processes, nonequilibrium conductance matrices exist beyond this framework: They can be employed to model a system directly without relying on microscopic dynamics.

\section{Serial association of thermodynamic devices}
\label{Serial Association}

\begin{figure}[t]
\includegraphics[width=\columnwidth]{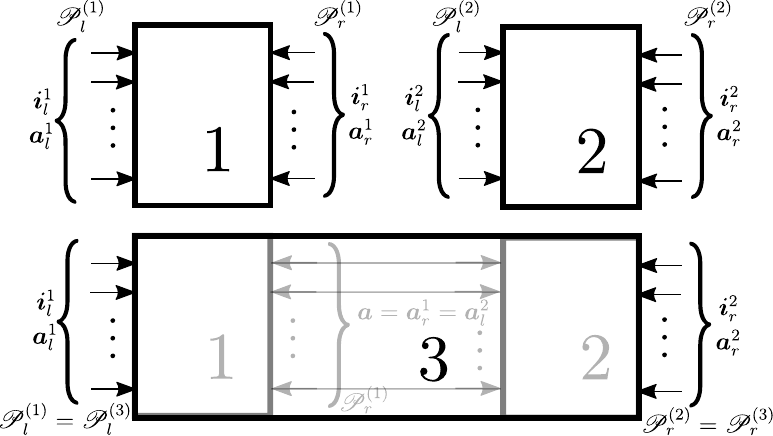}
\caption{(Top) Separated devices 1 and 2 with their various pins in the set $\mathscr{P}^{(m)}_{l}$ and $\mathscr{P}^{(m)}_{r}$. (Bottom) Merged device 3 made out of 1 and 2, with equal local potentials $\bm a$ on the wires connecting 1 and 2. Devices 1 and 2 are in mixed boundary conditions due to the pins in light grey (internal pins of device 3). The external pins in black are in Dirichlet conditions since the local potential on these pins is set by external reservoirs. \label{fig:device}}
\end{figure}

We now consider generic thermodynamic devices of given conductance matrix and conservation laws. In order to differentiate the devices in our thermodynamic circuit, we label them with an integer: a superscript $(m)$ is added to variables (local potentials, currents, conductance matrices, etc) to specify to which device they correspond. Boxes with several pins represent devices as in Fig.~\ref{fig:device}. A pin is a part of a device through which a current of a physical quantity can flow (e.g., thermal or electric contact). The pins of device $m$ are labeled with integers belonging to the sets $\mathscr{P}^{(m)}$. The pins on the left of device $m$ belong to  $\mathscr{P}^{(m)}_l$, those on the right to $\mathscr{P}^{(m)}_r$, such that $\mathscr{P}^{(m)}=\mathscr{P}^{(m)}_{l}\cup \mathscr{P}^{(m)}_{r}$ and $\mathscr{P}^{(m)}_{l}\cap \mathscr{P}^{(m)}_{r} = \emptyset  $. Without loss of generality, we connect the right pins of device 1 to the left pins of device 2 to create device $3$ as in Fig.~\ref{fig:device}. We use the same absolute labeling for the connected pins $\mathscr{P}^{(1)}_{r} = \mathscr{P}^{(2)}_{l} $ that we call internal pins from the viewpoint of device $3$. 
We proceed now with determining the conductance matrix at the fundamental level for the aforementioned serial association of devices. We start with the conductance matrices and conservation laws of each device. The serial association creates mixed boundary conditions on the connected side. In the first part below, we explain the method to determine the working point of the composite device. The currents conservation at the interface between the devices provides a set of equations leading to the stationary values of the local potentials on the connections. Then, those internal degrees of freedom are functions of the external local potentials imposed by reservoirs. In the second part, we provide the rule of resistance addition of the nonequilibrium conductance matrices for the serial association of two devices. Connecting more devices follows from a sequence of pairwise connections.

\subsection{Determination of internal degrees of freedom} 
The temperature, pressure, chemical potential, or any intensive variable at pin $p \in \mathscr{P}^{(m)}$ is the $p$th components $a^{(m)}_p$ of the local potential vector $\bm a^{(m)}$. Another device or a reservoir of the conjugated extensive quantity (e.g. heat, volume, chemical species, etc) sets the local potential at each pin. As in section \ref{NEcond}, the physical currents vector $\bm i^{(m)}$ has a positive $p$'th component $ i^{(m)}_p$ when a positive amount of the corresponding extensive quantity is received by device $m$ through pin $p$. We decompose the vectors of physical currents and local potentials into two sub-vectors as 
\begin{equation}
	\bm i^{(m)} = \begin{pmatrix}
	\bm i^{(m)}_l \\ \bm i^{(m)}_r
	\end{pmatrix}, \quad \text{ and } \quad 
	\bm a^{(m)} =  \begin{pmatrix} 
	\bm a_l^{(m)} \\ \bm a_r^{(m)} 
	\end{pmatrix}
\end{equation}
The sub-vector with index $\chi=l,r$ includes only the components on the $\chi$ side. By definition, we have 
\begin{equation}
\bm i^{(3)} = \begin{pmatrix} \bm i^{(1)}_l \\ \bm i^{(2)}_r \end{pmatrix},  \quad \text{ and } \quad 
	\bm a^{(3)} =  \begin{pmatrix} 
	\bm a_l^{(1)} \\ \bm a_r^{(2)} 
	\end{pmatrix}. \label{aANDiFORdevice3}
\end{equation} 
We exclude transient or periodic behavior. Our thermodynamic circuit has reached a nonequilibrium stationary state: for given $\bm a^{(m)}$, device $m$ has constant mean currents $\bm i^{(m)}$ that are nonlinear functions of $\bm a^{(m)}$.  
The stationary EPR for device $m$ reads $\sigma^{(m)}=\bm a^{(m)T}\bm i^{(m)}$. Assuming no dissipation at the interface, the EPR of device 3 is $\sigma^{(3)} = \sigma^{(1)}+\sigma^{(2)}$, leading to
\begin{eqnarray}
   \sigma^{(3)} &=& \bm a^{(1)T}_l  \bm i^{(1)}_l + \bm a^{(1)T}_r  \bm i^{(1)}_r + \bm a^{(2)T}_l  \bm i^{(2)}_l + \bm a^{(2)T}_r  \bm i^{(2)}_r, \nonumber \\
\end{eqnarray}
This is compatible with
\begin{eqnarray}
	  \sigma^{(3)}  &=& \bm a^{(3)T}\bm i^{(3)} = \bm a^{(1)T}_l  \bm i^{(1)}_l + \bm a^{(2)T}_r  \bm i^{(2)}_r , \label{additivity}
\end{eqnarray}
only if the local potentials on the connection pins are identical $\bm a^{(1)}_r = \bm a^{(2)}_l \equiv \bm a$ and given the current conservation at the interface
\begin{equation}
    \bm i^{(1)}_r + \bm i^{(2)}_l = 0.     \label{flux-conservation}
\end{equation}
Equation~\eqref{flux-conservation} is a nonlinear system of $|\mathscr{P}^{(1)}_r| = |\mathscr{P}^{(2)}_l| $ equations that must be solved to determine the internal local potentials. This provides $\bm a$ as a function of the external potentials $\bm a^{(1)}_l$ and $\bm a^{(2)}_r$ eliminating all internal degrees of freedom. 

\subsection{Equivalent conductance matrix at the fundamental level} 

Given the local potential $\bm a$ on the internal pins, devices $1$ and $2$ are driven out of equilibrium by local potentials 
\begin{equation}
	\bm a^{(1)} =  \begin{pmatrix} 
	\bm a_l^{(1)} \\ \bm a 
	\end{pmatrix}, \quad \text{ and } \quad 
	\bm a^{(2)} =  \begin{pmatrix} 
	\bm a \\ \bm a_r^{(2)} 
	\end{pmatrix}.
\end{equation}
Those are associated to fundamental forces following from Eq.~\eqref{FundForce}
\begin{equation}
	\bm A^{(1)} = \bm S^{(1)T} \bm a^{(1)}, \quad \text{ and } \quad \bm A^{(2)} = \bm S^{(2)T} \bm a^{(2)}.
\end{equation}
for some choice of selection matrix $\bm S^{(m)}$ associated to conservation laws $\bm \ell^{(m)}$ for $m=1,2$. Those forces lead to fundamental currents
\begin{equation}
	\bm I^{(1)} = \bm G^{(1)} \bm A^{(1)}, \quad \text{ and } \quad \bm I^{(2)} = \bm G^{(2)} \bm A^{(2)} 
\end{equation}
and to numerical values of the nonequilibrium conductance matrices $\bm G^{(m)}$ for $m=1,2$. We can now determine the rule of resistance addition for the inverses of those matrices. We do so by expressing the EPR of device $3$ as
\begin{equation}
	\sigma^{(3)} = \bm A^{(3)T} \bm I^{(3)} = \bm A^{(1)T} \bm I^{(1)} +\bm A^{(2)T} \bm I^{(2)} .
\end{equation}
For nonsingular conductance matrices, this yields
\begin{equation}
    \bm I^{(3)T}{\bm G^{(3)}}^{-1}\bm I^{(3)} = \sum_{m=1,2} \bm I^{(m)T}{\bm G^{(m)}} ^{-1}\bm I^{(m)}.
    \label{eq: EPR 3}
\end{equation}
Whenever the fundamental currents of devices $1$ and $2$ follow from those of device $3$, i.e. $\bm I^{(m)} =\bm{\Pi}^{(m,3)} \bm I^{(3)}$ for $m=1,2$, the non-equilibrium conductance matrix for device 3 reads
\begin{equation}
    {\bm G^{(3)}}^{-1}=\sum_{m=1}^2 \bm{\Pi}^{(m,3)T} {\bm G^{(m)}} ^{-1}\bm{\Pi}^{(m,3)}. \label{EquivImpedance}
\end{equation}
When the conductance matrices are nonsingular, the above equation generalizes the rule of resistance addition for the serial connection of thermodynamic devices. An explicit formula for matrix $\bm{\Pi}^{(m,3)}$ is provided in the next section when 
the devices' connection allows to integrate out the internal degrees of freedom. For more general connections, the system of Eq.~\eqref{flux-conservation} leading to the local potential on the connection pins is under-determined: Some local potentials are floating.

\subsection{Relation between fundamental currents}

As shown on Fig.~\ref{fig: exemple abstrait}, each conservation law $\bm \ell^{(m)}_{k}$ of devices $m=1,2$ can be represented as a tree graph, i.e. a graph connecting pins without making loops. More precisely, we represent each conservation law by a black square connected to several pins by lines associated with currents. With our convention of positive incoming currents, the sum of currents at each black square is zero. We emphasize that lines reaching a square represent current conservations only: Coupling between currents of different nature and being involved in different conservation laws are possible. As on Fig.~\ref{fig: exemple abstrait}, we assume that after connecting devices $1$ and $2$, the conservation laws of device $3$ are also tree graphs. If this was not the case, some potential would remain undetermined. For instance, let us consider the unique conservation law of device $1$ on Fig.~\ref{fig: exemple abstrait}. If we connect the pins $8$ and $9$ of device $2$ to pins $3$ and $4$ of device $1$, then two external reservoirs at pin $1$ and $2$ are insufficient to set the three fundamental forces driving device $1$ out of equilibrium. This trivial example illustrates the problem arising when making loops upon the connection of devices. In the presence of loops, the under-determination of the system can be overcome by adding a reservoir as an external gauge reference. This problem is similar to that of a floating mass in electrokinetics. Excluding connections creating loops, let us now determine the matrix $\bm{\Pi}^{(m,3)}$ relating the fundamental currents of device $m$ to those of device $3$. We proceed in two steps. First, we look for $\bm{\pi}^{(m,3)}$ satisfying $\bm i^{(m)} = \bm{\pi}^{(m,3)}\bm i^{(3)}$ since we have
\begin{equation}
    \bm{\Pi}^{(m,3)}  =\bm S^{(m)+} \bm \pi^{(m,3)}\bm S^{(3)},
    \label{eq: funda m to 3}
\end{equation}
when using Eqs.~(\ref{iSI}--\ref{S+iI}). Second, we determine the conservation laws of device $3$ leading to the selection matrix $\bm S^{(3)}$ required to get $\bm{\Pi}^{(m,3)}$ from $\bm{\pi}^{(m,3)}$.

First step: Let us relate physical currents of device $m$ and $3$. Since the physical currents vector decomposes into left and right sub-vectors, the matrix of conservation laws has two sub-matrices 
\begin{equation} 
\bm\ell^{(m)} \bm i^{(m)} = 
\left[\begin{array}{c|c}
    \bm\ell_l^{(m)} & \bm\ell_r^{(m)}
\end{array}\right]
\begin{pmatrix}
    \bm i_l^{(m)} \\
    \bm i_r^{(m)}
\end{pmatrix}
=\bm 0. 
\label{eq: split conservation laws}
\end{equation}
Using Eq.~\eqref{flux-conservation} for the current conservation at the interface yields
\begin{eqnarray}
	\bm\ell_l^{(1)} \bm i^{(1)}_{l}+ \bm\ell_r^{(1)} \bm i^{(1)}_{r} = 0, \label{device1} \\
	-\bm\ell_l^{(2)} \bm i^{(1)}_{r}+ \bm\ell_r^{(2)}\bm i^{(2)}_{r} = 0. \label{device2}
\end{eqnarray}
Introducing the matrices 
\begin{equation}
    \bm L_\mathrm{i} \equiv
    \begin{bmatrix}
        -\bm\ell_r^{(1)}\\
        \bm\ell_l^{(2)}
    \end{bmatrix}, \;
    \bm L_\mathrm{e} \equiv
    \begin{bmatrix}
        \bm\ell_l^{(1)} & \mathbb{0} \\
        \mathbb{0} & \bm\ell_r^{(2)}
    \end{bmatrix},
\end{equation}
Eqs.~(\ref{device1}--\ref{device2}) write
\begin{equation}
	\bm L_\mathrm{i} \bm i^{(1)}_r=\bm L_\mathrm{e} \bm i^{(3)}.     \label{eq: wire currents}
\end{equation}
Interestingly, the matrix $\bm L_\mathrm{i}$ is full column rank. In graph-theoretical language, $\bm L_\mathrm{i}$ is the incidence matrix of a tree graph whose vertices are the conservation laws and whose edges are the connection pins. We recall that an incidence matrix has lines associated with the graph's vertices and columns associated with edges connecting vertices. Each column contains zeros, but for the source vertex of the edge, that is $-1$, and for the target vertex, that is $+1$. The columns of an incidence matrix are linearly independent when the graph it represents has no cycle. In our case, the matrix $\bm L_\mathrm{i}$ summarizes the connection between conservation laws of device $1$ and $2$ via the internal pins in $ \mathscr{P}^{(1)}_{r} = \mathscr{P}^{(2)}_{l}$. Let us consider the $u$th and $v$th conservation laws of respectively devices 1 and 2 connected via the internal pin $p$. Then, the $p$th column of $\bm L_\mathrm{i}$ is made of zeros except at the lines of the connected conservation laws, i.e. when $( -\bm \ell^{(1)}_r)_{up} = -1 $ and $( \bm \ell^{(2)}_l )_{vp} = +1$, hence corresponding to the source and target of the connection via pin $p$. This shows that $\bm L_\mathrm{i}$ is an incidence matrix
that has independent columns from our assumption of pairwise connection of conservation laws producing tree graphs only (no loops). Therefore, we can define 
\begin{equation}
\bm{\pi} \equiv \bm L_\mathrm{i}^+\bm L_\mathrm{e} 
\end{equation}
and write 
\begin{equation}
	\bm i_r^{(1)}=\bm \pi \bm i^{(3)} = -\bm i_l^{(2)}.
\end{equation} 
We conclude this first step by introducing the matrices providing the physical current vector of device $m=1,2$ from the one of device 3:
\begin{equation}
    \bm{\pi}^{(1,3)} \equiv
    \begin{bmatrix}
        \begin{matrix}
            \mathbb{1} & \mathbb{0}
        \end{matrix}\\
        \hline
        \bm \pi
    \end{bmatrix}  \text{ and } 
    \bm{\pi}^{(2,3)}\equiv
    \begin{bmatrix}
        -\bm \pi\\
        \hline
        \begin{matrix}
            \mathbb{0} & \mathbb{1}
        \end{matrix}
\end{bmatrix},
\label{pimatrices}
\end{equation}
since $\bm i^{(1)}_l = [\mathbb{1}\; \mathbb{0}] \bm i^{(3)}$ and $\bm i^{(2)}_r = [ \mathbb{0}\; \mathbb{1} ] \bm i^{(3)}$ from Eq.~\eqref{aANDiFORdevice3}. 

Second step: Let us determine the matrix of conservation laws of device 3 from which the selection matrix $\bm S^{(3)}$ arises since its columns define a basis of $\mathrm{ker}\,(\bm \ell^{(3)})$. The number of conservation laws $|\mathscr{L}^{(3)}|$ is constrained by our assumption on the devices' connection. Indeed, there is a total of $|\mathscr{L}^{(1)}|+|\mathscr{L}^{(2)}|$ conservation laws for the disconnected devices. Each internal pin connects two conservation laws reducing this total by one and leading to
\begin{equation}
	|\mathscr{L}^{(3)}|=|\mathscr{L}^{(1)}|+|\mathscr{L}^{(2)}|-| \mathscr{P}_r^{(1)}|.  \label{cardL3}
\end{equation}
On the other side, we denote by $\bm v$ the matrix satisfying 
\begin{equation}
\bm v \bm L_\mathrm{i}  = 0 .
\end{equation}
The line vectors of $\bm v$ provide a basis of $\bm L_\mathrm{i}$'s cokernel. The rank-nullity theorem applied to $\bm L_\mathrm{i}^{T}$ states that
\begin{equation}
    \mathrm{dim}\,\left( \mathrm{ker}\, \bm L_\mathrm{i}^{T} \right) +\mathrm{rk}\, (\bm L_\mathrm{i}^{T}) =|\mathscr{L}^{(1)}|+|\mathscr{L}^{(2)}|.
\end{equation} 
Since $\bm L_\mathrm{i}$ has independent columns, we have
\begin{equation}
	\mathrm{rk}\, (\bm L_\mathrm{i}^{T}) = \mathrm{rk}\, (\bm L_\mathrm{i}) = |\mathscr{P}_r^{(1)}|.
\end{equation}
Therefore, $|\mathscr{L}^{(3)}|$ of Eq.~\eqref{cardL3} coincides with the number of lines of matrix $\bm v$. We conclude that all the conservation laws of device $3$ follow from left multiplying Eq.~\eqref{eq: wire currents} by matrix $\bm v$: The matrix of conservation laws reads $\bm \ell^{(3)} = \bm v \bm L_\mathrm{e} $ and has the correct number of lines. A basis of its kernel leads to $\bm S^{(3)}$, with a choice of fundamental forces and currents that is free in our theory. This ends our demonstration that it is possible to write the physical and fundamental currents of device $m=1,2$ using those of device $3$ according to Eq.~\eqref{eq: funda m to 3} providing $\bm{\Pi}^{(m,3)}$ via Eq.~\eqref{pimatrices} and the chosen matrix $\bm S^{(3)}$.

\subsection{Illustration of the serial association}

\begin{figure}[b]
\includegraphics[width=\columnwidth]{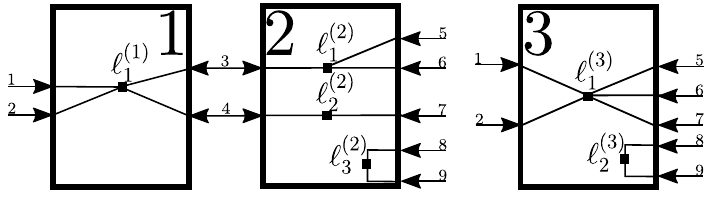}
\caption{Example of the serial connection of two devices.
For device $m=1$ the pins are divided into $\mathscr{P}^{(1)}_l= \{1,2\}$ and $\mathscr{P}^{(1)}_r= \{3,4\}$. For device $m=2$ the pins are divided into $\mathscr{P}^{(2)}_l= \{3,4\}$ and $\mathscr{P}^{(2)}_r= \{5,6,7,8,9\}$.  \label{fig: exemple abstrait}}
\end{figure}

 We now illustrate the construction of an equivalent nonequilibrium conductance for the devices of Fig.~\ref{fig: exemple abstrait}. 
The first and second devices have 4 and 7 pins, 1 and 3 conservation laws leading to 3 and 4 fundamental currents respectively. Their connection via $\bm i_r^{(1)}=\left(i_3, i_4 \right)^T$ produces a third device with 7 pins, 2 conservation laws and 5 fundamental currents. 
Hence, none of the conductance matrices are of the same dimension: this illustrates the most general serial association of devices. We assume that the local potentials at the interface have already been determined and that the conductance matrix $\bm G^{(m)}$ for $m=1,2$ are known for fundamental currents $\bm I^{(1)}=(i_2,i_3,i_4)^T$ and $\bm I^{(2)}=(i_5,i_6,i_7,i_9)^T$ associated to 
\begin{equation}
    \bm S^{(1)}=
    \begin{bmatrix}
        -1 & -1 & -1 \\
        1  &  0 &  0 \\
        0  &  1 &  0 \\
        0  &  0 &  1
    \end{bmatrix}, \quad
    \bm S^{(2)}=
    \begin{bmatrix}
        -1 & -1 & 0 & 0 \\
        0  &  0 & -1 & 0\\
        1 & 0 & 0 & 0 \\
        0 & 1 & 0 & 0 \\
        0 & 0 & 1 & 0 \\
        0 & 0 & 0 & -1\\
        0 & 0 & 0 & 1
    \end{bmatrix}. \label{SelectionMatrix12}
\end{equation}
Eq.~\eqref{iSI} leads to physical currents $\bm i^{(1)}=(i_1,\dots,i_4)^T$ and $\bm i^{(2)}=(i_3,\dots,i_9)^T$ as expected given the matrices 
\begin{equation}
    \bm \ell^{(1)}=
    \left[\begin{array}{cc|cc}
        1 & 1 & 1 & 1
    \end{array}\right],
\quad 
    \bm \ell^{(2)}=
    \left[\begin{array}{cc|ccccc}
        1 & 0 & 1 & 1 & 0 & 0 & 0 \\
        0 & 1 & 0 & 0 & 1 & 0 & 0 \\
        0 & 0 & 0 & 0 & 0 & 1 & 1
    \end{array}\right],
\end{equation}
for the conservation laws $\bm \ell^{(m)} \bm i^{(m)} = \bm 0$ depicted on Fig.~\ref{fig: exemple abstrait}. Vertical bars emphasize the bloc decomposition of Eq.~\eqref{eq: split conservation laws}. Applying Eq.~\eqref{eq: wire currents} to our example, we find
\begin{equation}
    \bm L_\mathrm{e} = 
    \begin{array}{c}
         \begin{array}{ccccccc}
         1&2&5&6&7&8&9
         \end{array}
         \\        
    \left[\begin{array}{cc|ccccc}
        1 & 1 & 0 & 0 & 0 & 0 & 0 \\ \hline
        0 & 0 & 1 & 1 & 0 & 0 & 0 \\
        0 & 0 & 0 & 0 & 1 & 0 & 0 \\
        0 & 0 & 0 & 0 & 0 & 1 & 1
    \end{array}\right]
    \end{array},
    \quad 
    \bm L_\mathrm{i} =
    \begin{array}{c}
     \begin{array}{cc} 
     3&\;\;4
     \end{array}
     \\
    \left[\begin{array}{cc}
        -1 & -1 \\ \hline
        1 & 0 \\
        0 & 1 \\
        0 & 0 
    \end{array}\right]
    \end{array},
\end{equation}
where we indicate the pin index at the top of each column. 
As expected, the matrix $\bm L_\mathrm{i}$ is full column rank with pseudo  inverse 
\begin{equation}
    \bm L_\mathrm{i}^+=\frac{1}{3} 
    \begin{pmatrix}
        -1 & 2 & -1 & 0 \\
        -1 & -1 & 2 & 0
    \end{pmatrix}.
\end{equation}
Matrices $\bm \pi^{(m,3)}$ for $m=1,2$ of Eq.~\eqref{pimatrices} read
\begin{eqnarray}
\bm\pi^{(1,3)}&=&
    \left[
    \def\arraystretch{1.3}
    \begin{array}{C{1.5em}C{1.5em}C{1.5em}C{1.5em}C{1.5em}C{1.5em}C{1.5em}}
				1 & 0 & 0&0&0&0&0 \\
				0 & 1 & 0&0&0&0&0 \\
        \hline
            -\frac{1}{3} & -\frac{1}{3} & \frac{2}{3} & \frac{2}{3} & -\frac{1}{3} & 0 & 0 \\
            -\frac{1}{3} & -\frac{1}{3} & -\frac{1}{3} & -\frac{1}{3} & \frac{2}{3} & 0 & 0   
    \end{array}
    \right], \label{pi1} \\
\bm\pi^{(2,3)} &=& 
    \left[
    \def\arraystretch{1.3}
    \begin{array}{C{1.5em}C{1.5em}C{1.5em}C{1.5em}C{1.5em}C{1.5em}C{1.5em}}
            \frac{1}{3} & \frac{1}{3} & -\frac{2}{3} & -\frac{2}{3} & \frac{1}{3} & 0 & 0 \\
            \frac{1}{3} & \frac{1}{3} & \frac{1}{3} & \frac{1}{3} & -\frac{2}{3} & 0 & 0 \\
        \hline
        0 & 0 &   1 & 0& 0& 0& 0  \\
        0 & 0 &   0 & 1& 0& 0& 0  \\
        0 & 0 &   0 & 0& 1& 0& 0  \\
        0 & 0 &   0 & 0& 0& 1& 0  \\
        0 & 0 &   0 & 0& 0& 0& 1  
    \end{array}
    \right]. \label{pi2}
\end{eqnarray}
Next, the choice of selection matrix $\bm S^{(3)}$ determines the fundamental basis in which $\bm G^{(3)}$ is given. The column vectors of $\bm S^{(3)}$ realize a basis of $\mathrm{ker}(\bm\ell^{(3)})$. Let us first determine $\bm\ell^{(3)}$ by finding the left null eigenvectors of $\bm L_\mathrm{i}$ that we gather in the lines of matrix
\begin{equation}
    \bm v = \begin{bmatrix}
        1 & 1 & 1 & 0 \\
        0 & 0 & 0 & 1
    \end{bmatrix}.
\end{equation}
Then, the matrix of conservation laws arises from
\begin{equation}
    \bm \ell^{(3)}=\bm v \bm L_\mathrm{e} = 
    \begin{bmatrix}
        1 & 1 & 1 & 1 & 1 & 0 & 0 \\
        0 & 0 & 0 & 0 & 0 & 1 & 1
    \end{bmatrix}.
\end{equation}
The vector of physical currents $\bm i^{(3)}=(i_1,i_2,i_5,i_6,i_7,i_8,i_9)^T$
is obtained from the product of the selection matrix
\begin{equation}
    \bm S^{(3)} = 
    \begin{bmatrix}
        -1 & -1 & -1 & -1 & 0 \\
        1 & 0 & 0 & 0 & 0 \\
        0 & 1 & 0 & 0 & 0 \\
        0 & 0 & 1 & 0 & 0 \\
        0 & 0 & 0 & 1 & 0 \\
        0 & 0 & 0 & 0 & -1\\
        0 & 0 & 0 & 0 & 1
    \end{bmatrix} \label{S3}
\end{equation}
with the vector of fundamental currents $\bm I^{(3)}=\left( i_2, i_5, i_6, i_7, i_9 \right)^T$. Finally, the matrices $\bm \Pi^{(m,3)}$ for $m=1,2$ and then $\bm G^{(3)}$ follow from Eqs.~(\ref{EquivImpedance}--\ref{eq: funda m to 3}) with the left pseudoinverses of $\bm S^{(m)}$ in Eq.~\eqref{SelectionMatrix12}, the $\bm{\pi}^{(m,3)}$ matrices of Eqs.~(\ref{pi1}--\ref{pi2}) and the selection matrix of Eq.~\eqref{S3}.

\section{Parallel association of thermodynamic devices}
\label{Parallel Association}
Given their conductances and conservation laws, we continue with the determination of the conductance matrix at the physical and fundamental levels for a composite device made of two sub-devices in parallel association. Two devices are connected in parallel if at least a first reservoir imposes a local potential on a first set of pins (including pin(s) of device 1 and 2) and a second reservoir does the same for a second set of pins (including other pin(s) of device 1 and 2). Then, a physical current flows from one reservoir to another across the two devices in parallel. If this is not the case, the two sets of pins are not connected in parallel, although this situation may also be of some interest. The framework of this section can deal with all these situations. In electrokinetics, the parallel connection of two dipoles produces another dipole, not a tripole. Indeed, a scalar resistance (or conductance) describes only dipoles. Our framework is more general since conductance matrices describe multipoles. 

As for the serial connection, the set of pins of device $m$ is $\mathscr{P}^{(m)}$. For $m=1,2$, these sets include integers generically denoted $p$. Contrarily to serial connection, the set $\mathscr{P}^{(1)}$ and $\mathscr{P}^{(2)}$ are disjoint: their intersection does not give the connected pins anymore. This allows to determine $m$ from the value of $p\in \mathscr{P}^{(1)}\cup \mathscr{P}^{(2)}$. Still contrarily to serial connection, the elements of $\mathscr{P}^{(3)}$ are sets of integers generically denoted $P$. We call $P$ the lumped pins: they correspond to pins connected to the same reservoir to impose an identical local potential, i.e., $\forall \, p \in P$, $a^{(m)}_p = a^{(3)}_P$. From now on, we drop the device's superscript on components of vectors: the pin index ($p$ or $P$) with its letter case is sufficient to determine the considered device $m \in \{1,2,3\}$. We remark that $\forall \, p \in \mathscr{P}^{(1)}\cup \mathscr{P}^{(2)}, \exists ! P \in \mathscr{P}^{(3)} | p \in P$. In other words, the intersection of any two lumped pins is empty. The union of lumped pins gives the complete set of pins
\begin{equation}
    \bigcup_{P\in \mathscr{P}^{(3)}} P = \mathscr{P}^{(1)}\cup \mathscr{P}^{(2)}.
\end{equation}
Since device 3 remains in Dirichlet boundary conditions, there is no need to determine the working point of the devices, as we did for serial connection. We can compute directly the nonequilibrium conductance matrix of device 3 from those of devices 1 and 2 in our parallel setting. We do so at the level of physical currents and local potentials. Hence, we need to determine the selection matrix for device 3 to provide the nonequilibrium conductance matrix at the fundamental level. Finally, we illustrate the results of this section by connecting in parallel devices 1 and 2 of the previous section.

\subsection{Equivalent conductance matrix at the physical level}
This section provides the equivalent conductance matrix for the parallel connection of two devices. 
As for a serial connection, the \emph{EPR} of device 3 reads:
\begin{equation}
    \sigma^{(3)}
    =\sum_{m=1}^2\bm {a}^{(m)T}\bm i^{(m)}=\sum_{m=1}^2\bm a^{(m)T}\bm g^{(m)}\bm a^{(m)}.
    \label{eq : EPR parallel}
\end{equation}
By definition of a parallel connection, the local potentials within lumped pins take the same value. This means that the local potentials at all pins of devices 1 and 2 are functions of those on the lumped pin of device 3, according to
\begin{equation}
    \bm a^{(m)}=\bm \pi_{\parallel}^{(m,3)}\bm a^{(3)},
    \label{eq : passage am to a3}
\end{equation}
where $\bm \pi_{\parallel}^{(m,3)}$ is a rectangular matrix of dimension $|\mathscr{P}^{(m)}| \times |\mathscr{P}^{(3)}|$ defining how pins are lumped together in the parallel connection
\begin{equation}
    \left(\bm \pi_{\parallel}^{(m,3)}\right)_{p,P}=\left\{ 
  \begin{array}{ c l }
    1 & \quad \text{if $p\in P$ for $P \in \mathscr{P}^{(3)}$}, \\
    0                 & \quad \text{ otherwise.}
  \end{array}
\right.
\label{eq : matrice indicatrice}
\end{equation}
Using Eq.~\eqref{eq : passage am to a3} in Eq.~\eqref{eq : EPR parallel} yields
\begin{equation}
    \sigma^{(3)}=\bm a^{(3)T}\left( \sum_{m=1}^2  \bm\pi_{\parallel}^{(m,3)T}\bm g^{(m)}\bm\pi_{\parallel}^{(m,3)} \right)\bm a^{(3)}.
    \label{eq : EPR expression 1}
\end{equation}
Given that the \emph{EPR} of device 3 must read 
\begin{equation}
    \sigma^{(3)}=\bm a^{(3)T}\bm g^{(3)}\bm a^{(3)},
    \label{eq : EPR expression 2}
\end{equation}
we identify the conductance matrix at the level of physical currents and local potentials 
\begin{equation}
    \bm g^{(3)}=\sum_{m=1}^2  \bm\pi_{\parallel}^{(m,3)T}\bm g^{(m)}\bm\pi_{\parallel}^{(m,3)}.
    \label{eq : conductance parallel}
\end{equation}
The conductance matrix at the level of fundamental currents and forces arises from choosing a selection matrix associated with the conservation laws of device 3 that we determine in the next section.
\subsection{Conservation laws}

Let us determine the conservation laws for device 3 made out of devices 1 and 2 in parallel. We know the conservation laws $\bm l^{(m)}$ for $m=1,2$ for the two subdevices. We call them internal conservation laws from the point of view of device 3. In addition, the parallel connection imposes ``new'' conservation laws between the pins of the subdevices and the lumped pins of device 3. We call these new conservation laws external, and we gather them in the following relation:
\begin{equation}
\bm i^{(3)}=
\begin{bmatrix}
\bm \pi_\parallel^{(1,3)T} & \bm \pi_\parallel^{(2,3)T}
\end{bmatrix}
\begin{pmatrix}
\bm i^{(1)}\\
\bm i^{(2)}
\end{pmatrix}
\end{equation}
We gather all the conservation laws in the following relation:
\begin{equation}
\bm L
\begin{pmatrix}
\bm i^{(1)}\\
\bm i^{(2)}
\end{pmatrix}=
\begin{pmatrix}
\bm 0\\
\bm i^{(3)}
\end{pmatrix}
\label{eq : conservation laws}
\end{equation}
where:
\begin{equation}
\bm L=\begin{bmatrix}
\bm l^{(1)} & \bm 0\\
0 & \bm l^{(2)}\\
\bm \pi_\parallel^{(1,3)T} & \bm \pi_\parallel^{(2,3)T}
\end{bmatrix}. \label{defL}
\end{equation}
We show below that the cokernel of $\bm L$ has dimension $|\mathscr{L}^{(3)}|$. This provides a way to determine the conservation laws of device $3$ by a simple eigenvalue problem. More precisely, each null left eigenvector of $\bm L$ has its last $|\mathscr{P}^{(3)}|$ components giving a conservation law vector $\bm {\ell}^{(3)}_{k}$. As a first step of our proof, we study the structure of matrix $\bm L$. Then, in a second step, we use this structure to determine the conservation laws $\bm l^{(3)}$ of device $3$. The next section provides an example that usefully illustrates this proof. It can thus be read in parallel.

First, the matrix $\bm L$ has $ \vert \mathscr{L}^{(1)}\vert + \vert \mathscr{L}^{(2)}\vert + \vert \mathscr{P}^{(3)}\vert$ rows and $\vert \mathscr{P}^{(1)}\vert + \vert \mathscr{P}^{(2)}\vert$ columns. It describes the parallel association of devices $1$ and $2$. 
Each column of matrix $\bm L$, labeled by $p$, includes exactly two components equal to $+1$ while all other components are null: The first $\vert \mathscr{L}^{(1)}\vert + \vert \mathscr{L}^{(2)}\vert$ lines of  column $p$ includes just one $+1$ component because the pin current $i_{p}$ cannot appear in more than one conservation law. Similarly, the last $\vert \mathscr{P}^{(3)}\vert$ components of column $p$ include another $+1$, given that the lumped pins $P$ are disjoint sets whose union includes all pins of device $1$ and $2$.
With such columns, the matrix $\bm L$ is an incidence matrix of an undirected graph. The graph vertices are the conservation laws (both internal and external), and the edges are the pins of devices $1$ and $2$ (before associating the device in parallel). The graph associated with matrix $\bm L$ indicates how internal and external conservation laws are connected through the lumping of the pins. Finally, for an open system, the internal conservation laws for energy and matter cannot be connected by external conservation laws. Hence, the graph associated with the incidence matrix $\bm L$ has two disconnected components at least. Therefore, given the structure described above, it can be reduced to a block diagonal form \emph{by lines permutations only}, with one block for each disconnected component in the graph associated with $\bm L$. We remark that the lines permutation uses the freedom in labeling the internal and external conservation laws. We also remark that we do not need column permutations thanks to our numbering convention for pins: physical currents appearing in the same conservation law are associated with pins numbered with consecutive integers.
We denote by $\tilde{\bm L}$ the obtained block diagonal matrix such that
\begin{equation}
\tilde{\bm L}=\bigoplus_{\textsc{c}} \tilde{\bm L}_{\textsc{c}} \label{BlockForm}
\end{equation}
where each block matrix $\tilde{\bm L}_{\textsc{c}}$ is the incidence matrix for the connected component $\textsc{c}$ of the graph of conservation laws. Since the incidence matrix of a connected graph admits a single left null vector \cite{Avanzini2024vol}, the dimension of $\mathrm{coker}(\bm L)=\mathrm{coker}( \tilde{\bm L})$ is equal to the number of connected components in the graph. By construction, this number is equal to the number of conservation laws $\vert \mathscr{L}^{(3)}\vert$ for device $3$. 

In a second step, let us obtain the conservation laws for device 3 such that
$ \bm l^{(3)}\bm i^{(3)}=\bm 0$. The lines reordering leading to the block diagonal incidence matrix of Eq.~\eqref{BlockForm} implies reordering the currents components on the right-hand side of Eq.~\eqref{eq : conservation laws}. Since the order of lines in each block diagonal part $\tilde{\bm L}_{\textsc{c}}$ is arbitrary, we take by convention that the first $\vert \mathscr{L}^{(1)}_{\textsc{c}} \vert $ lines are for the internal conservation law of device 1 appearing in $\textsc{c}$. The following $\vert \mathscr{L}^{(2)}_{\textsc{c}} \vert $ lines are those of device 2 appearing in $\textsc{c}$. The last $\vert \mathscr{P}^{(3)}_{\textsc{c}}\vert$ lines are the external conservation laws, with one line per lumped pin appearing in the connected graph described by $\tilde{\bm L}_{\textsc{c}}$. With this convention, each matrix $\tilde{\bm L}_{\textsc{c}}$ is very close in definition to matrix $\bm L$ as defined in Eq.~\eqref{defL}, except that it describes just the connected component $\textsc{c}$. Then, Eq.~\eqref{eq : conservation laws} reads block-wise
\begin{equation}
 \tilde{\bm L}_{\textsc{c}} \tilde {\bm {i}}_\textsc{c}= \begin{pmatrix}
\bm 0\\
\tilde{\bm i}^{(3)}_{\textsc{c}}
\end{pmatrix} , \label{eq : conservation laws c}
\end{equation}
where $\tilde {\bm i}_{\textsc{c}}$ is the vector of physical currents going through the sub-devices pins belonging to the connected component $\textsc{c}$, similarly $\tilde{\bm i}^{(3)}_{\textsc{c}}$ is the vector of physical currents going through the lumped pins of the same connected component. The later column vector has dimension $\vert \mathscr{P}^{(3)}_{\textsc{c}} \vert$, while $\bm 0$ is the null column vector of dimension $\vert \mathscr{L}^{(1)}_{\textsc{c}} \vert + \vert \mathscr{L}^{(2)}_{\textsc{c}} \vert$. Since we used line permutations only, we remark that the columns' order in matrix $\tilde {\bm L}$ is identical to that of $\bm L$. Then, the order of the physical currents is preserved, and we have the identity
\begin{equation}
	\begin{pmatrix}
\bm i^{(1)}\\
\bm i^{(2)}
\end{pmatrix} = \begin{pmatrix} \tilde{\bm i}_\textsc{c1} \\ \tilde{\bm i}_\textsc{c2} \\ \vdots \end{pmatrix}.
\end{equation}
The tilde vectors $\tilde {\bm{i}}_\textsc{ci}$ extract different subsets of physical currents still appearing in the same order. 
Given that the matrix $\bm L_\textsc{c}$ are incidence matrices of a connected undirected graph, the row vector $\bm u_{\textsc{c}}$ in $\mathrm{coker}(\tilde{\bm L}_{\textsc{c}})$ writes
\begin{equation}
\bm u_{\textsc{c}}=
\begin{pmatrix}
-1 \hdots -1 & 1 \hdots 1
\end{pmatrix}
\end{equation}
with the first $\vert \mathscr{L}^{(1)}_c \vert + \vert \mathscr{L}^{(2)}_c \vert$ components equal to $-1$ and the last $\vert \mathscr{P}^{(3)}_c\vert$ components equal to $1$.
Using $\tilde{\bm L}_{\textsc{c}} \bm u_{\textsc{c}} = 0$ we find
\begin{equation}
\bm u_{\textsc{c}} \tilde{\bm L}_{\textsc{c}} \tilde{\bm i}_{\textsc{c}}= 
0= \begin{pmatrix}
-1 \hdots -1 & 1 \hdots 1
\end{pmatrix} \begin{pmatrix}
\bm 0\\
\tilde{\bm i}^{(3)}_{\textsc{c}}
\end{pmatrix} .
\end{equation}
We have found one conservation law per connected component $\textsc{c}$: the sum of all currents entering device 3 by the lumped pins of the connected component $\textsc{c}$ is null. In block diagonal form, this gives the matrix of conservation law
\begin{equation}
\bm \ell^{(3)}=\bigoplus_\textsc{c}
\bm \ell^{(3)}_\textsc{c}
\end{equation}
where $\bm \ell^{(3)}_\textsc{c}$ are row vectors with $|\mathscr{P}_\textsc{c}^{(3)}|$ components equal to $1$. 
In the next section, we illustrate this block diagonal decomposition of the matrix of conservation laws for device 3. Beyond this decomposition, we conclude that looking directly at the last components of the vectors in the cokernel of matrix $\bm L$ directly produces the conservation laws of device 3. This systematic approach is thus useful for large circuits with many currents and conservation laws.

\subsection{Illustration of the parallel association}
\begin{figure}[t] \centering
\includegraphics[width=\columnwidth]{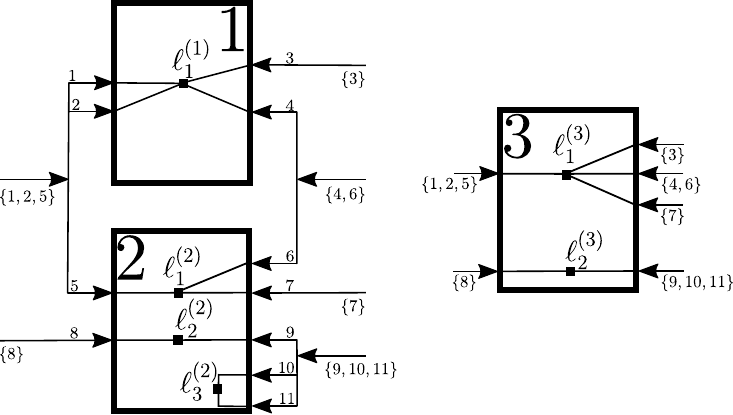}
\caption{(Left) Parallel connection of devices 1 and 2. The set of pins $\{1,2,5\}$ are connected to the same reservoir (of the appropriate physical quantity) that imposes the identical local potential $a^{(3)}_{\{1,2,5\}}$ on these pins. Similarly, the local potentials for the sets $\{4,6\}$ and $\{9,10,11\}$ are $a^{(3)}_{\{4,6\}}$ and 
$a^{(3)}_{\{9,10,11\}}$ respectively. (Right) Device 3 resulting from the parallel connection of devices 1 and 2 with its two conservation laws. \label{fig:parallelconnection}}
\end{figure}
In this section, devices 1 and 2 already introduced for the serial connection of Fig.~\ref{fig: exemple abstrait} are now connected in parallel according to Fig.~\ref{fig:parallelconnection}.
The pin ensembles are
\begin{eqnarray}
\mathscr{P}^{(1)}&=&\lbrace 1,2,3,4 \rbrace,\\
\mathscr{P}^{(2)}&=&\lbrace 5,6,7,8,9,10,11 \rbrace,\\
\mathscr{P}^{(3)}&=&\lbrace \lbrace 1,2,5\rbrace,\lbrace 3\rbrace,\lbrace 4,6\rbrace,\lbrace 7\rbrace,\lbrace 8\rbrace,\lbrace 9,10,11\rbrace \rbrace . \nonumber \\ \label{eq : P3}
\end{eqnarray}

The local potentials on the pins of machine $m=1,2$ are related to the local potential on the pins of machine $3$ by Eq.~\eqref{eq : passage am to a3} with
\begin{eqnarray}
    \bm\pi_{\parallel}^{(1,3)}&=&
    \begin{bmatrix}
        1 & 0 & 0 & 0 & 0 & 0\\
        1 & 0 & 0 & 0 & 0 & 0\\
        0 & 1 & 0 & 0 & 0 & 0\\
        0 & 0 & 1 & 0 & 0 & 0\\
    \end{bmatrix} \quad \begin{matrix} 1\\2 \\3 \\4 \end{matrix}, 
    \\
        \bm\pi_{\parallel}^{(2,3)}&=&
    \begin{bmatrix}
        1 & 0 & 0 & 0 & 0 & 0\\
        0 & 0 & 1 & 0 & 0 & 0\\
        0 & 0 & 0 & 1 & 0 & 0\\
        0 & 0 & 0 & 0 & 1 & 0\\
        0 & 0 & 0 & 0 & 0 & 1\\
        0 & 0 & 0 & 0 & 0 & 1\\
        0 & 0 & 0 & 0 & 0 & 1
    \end{bmatrix}\quad \begin{matrix} 5\\6 \\7 \\8 \\9 \\10 \\11 \end{matrix},
    \label{eq : pi parallel 1,2}
\end{eqnarray}
where the rows of the matrices are numbered on the right with the pin indices of device $m=1,2$, and the columns with the pin lumped indices of device 3 ordered as in Eq.~\eqref{eq : P3}.
The above matrices lead to the conductance matrix of device 3 using Eq.~\eqref{eq : conductance parallel}.
The conservation laws of devices 1 and 2 are listed in the following matrices:
\begin{equation}
    \bm \ell^{(1)}=\begin{bmatrix}
        1 & 1 & 1 & 1
    \end{bmatrix},\quad \bm \ell^{(2)}=\begin{bmatrix}
        1 & 1 & 1 & 0 & 0 & 0 & 0\\
        0 & 0 & 0 & 1 & 1 & 0 & 0\\
        0 & 0 & 0 & 0 & 0 & 1 & 1\\
    \end{bmatrix}.
\end{equation}
The matrix $\bm L$ thus reads in this case
\begin{equation}
\bm L=
\left[
\begin{array}{ccccccccccc}
 1 & 1 & 1 & 1 & 0 & 0 & 0 & 0 & 0 & 0 & 0
   \\
 0 & 0 & 0 & 0 & 1 & 1 & 1 & 0 & 0 & 0 & 0
   \\
 0 & 0 & 0 & 0 & 0 & 0 & 0 & 1 & 1 & 0 & 0
   \\
 0 & 0 & 0 & 0 & 0 & 0 & 0 & 0 & 0 & 1 & 1
   \\
 1 & 1 & 0 & 0 & 1 & 0 & 0 & 0 & 0 & 0 & 0
   \\
 0 & 0 & 1 & 0 & 0 & 0 & 0 & 0 & 0 & 0 & 0
   \\
 0 & 0 & 0 & 1 & 0 & 1 & 0 & 0 & 0 & 0 & 0
   \\
 0 & 0 & 0 & 0 & 0 & 0 & 1 & 0 & 0 & 0 & 0
   \\
 0 & 0 & 0 & 0 & 0 & 0 & 0 & 1 & 0 & 0 & 0
   \\
 0 & 0 & 0 & 0 & 0 & 0 & 0 & 0 & 1 & 1 & 1
   \\
\end{array}
\right],
\end{equation}
and the associated current vectors are
\begin{equation}
\begin{pmatrix}
\bm i^{(1)}\\
\bm i^{(2)}
\end{pmatrix}=
\begin{pmatrix}
i_1\\
i_2\\
i_3\\
i_4\\
i_5\\
i_6\\
i_7\\
i_8\\
i_9\\
i_{10}\\
i_{11}
\end{pmatrix},\quad
\begin{pmatrix}
\bm 0\\
\bm i^{(3)}
\end{pmatrix}=
\begin{pmatrix}
0 \\
0 \\
0 \\
0 \\
i_{\lbrace 1,2,5\rbrace}\\
i_{\lbrace 3 \rbrace}\\
i_{\lbrace 4,6\rbrace}\\
i_{\lbrace 7\rbrace}\\
i_{\lbrace 8\rbrace}\\
i_{\lbrace 9,10,11\rbrace}
\end{pmatrix}.
\end{equation}
After moving the third and fourth lines of $\bm L$ just before the last two lines, we reach the block diagonal form
\begin{equation}
\tilde{\bm L}=
\left[
\begin{array}{ccccccc|cccc}
 1 & 1 & 1 & 1 & 0 & 0 & 0 & 0 & 0 & 0 & 0
   \\
 0 & 0 & 0 & 0 & 1 & 1 & 1 & 0 & 0 & 0 & 0
   \\
 1 & 1 & 0 & 0 & 1 & 0 & 0 & 0 & 0 & 0 & 0
   \\
 0 & 0 & 1 & 0 & 0 & 0 & 0 & 0 & 0 & 0 & 0
   \\
 0 & 0 & 0 & 1 & 0 & 1 & 0 & 0 & 0 & 0 & 0
   \\
 0 & 0 & 0 & 0 & 0 & 0 & 1 & 0 & 0 & 0 & 0
 \\ \hline
 0 & 0 & 0 & 0 & 0 & 0 & 0 & 1 & 1 & 0 & 0
   \\
 0 & 0 & 0 & 0 & 0 & 0 & 0 & 0 & 0 & 1 & 1
   \\
 0 & 0 & 0 & 0 & 0 & 0 & 0 & 1 & 0 & 0 & 0
   \\
 0 & 0 & 0 & 0 & 0 & 0 & 0 & 0 & 1 & 1 & 1
   \\
\end{array}
\right].
\end{equation} 
In our case, the graph connecting conservation laws to give device 3 has  two disconnected components since $\tilde{\bm L}$ has two blocks $\tilde{\bm L}_{\textsc{c}1}$ and $\tilde{\bm L}_{\textsc{c}2}$ on its diagonal.
These blocks read
\begin{equation}
\tilde{\bm L}_{\textsc{c}1}=
\left[
\begin{array}{ccccccccccc}
 1 & 1 & 1 & 1 & 0 & 0 & 0 
   \\
 0 & 0 & 0 & 0 & 1 & 1 & 1 
   \\
 1 & 1 & 0 & 0 & 1 & 0 & 0 
   \\
 0 & 0 & 1 & 0 & 0 & 0 & 0 
   \\
 0 & 0 & 0 & 1 & 0 & 1 & 0 
   \\
 0 & 0 & 0 & 0 & 0 & 0 & 1 
\end{array}
\right],\quad
\tilde{\bm L}_{\textsc{c}2}=
\left[
\begin{array}{ccccccccccc}
 1 & 1 & 0 & 0
   \\
 0 & 0 & 1 & 1
   \\
 1 & 0 & 0 & 0
   \\
 0 & 1 & 1 & 1
   \\
\end{array}
\right]
\end{equation} 
and their associated current vectors are respectively
\begin{equation}
\tilde{\bm i}_{\textsc{c}1}=
\begin{pmatrix}
i_1\\
i_2\\
i_3\\
i_4\\
i_5\\
i_6\\
i_7
\end{pmatrix},\quad
\begin{pmatrix}
\bm 0\\
\tilde{\bm i}^{(3)}_{\textsc{c}1}
\end{pmatrix}=
\begin{pmatrix}
0 \\
0 \\
i_{\lbrace 1,2,5\rbrace}\\
i_{\lbrace 3 \rbrace}\\
i_{\lbrace 4,6\rbrace}\\
i_{\lbrace 7\rbrace}
\end{pmatrix}
\end{equation}
and 
\begin{equation}
\tilde{\bm i}_{\textsc{c}2}=
\begin{pmatrix}
i_8\\
i_9\\
i_{10}\\
i_{11}
\end{pmatrix},\quad
\begin{pmatrix}
\bm 0\\
\tilde{\bm i}^{(3)}_{\textsc{c}2}
\end{pmatrix}=
\begin{pmatrix}
0\\
0\\
i_{\lbrace 8\rbrace}\\
i_{\lbrace 9,10,11\rbrace}
\end{pmatrix}
\end{equation}
One can quickly check that the following vectors are in the left null spaces of respectively $\tilde{\bm L}_{\textsc{c}1}$ and $\tilde{\bm L}_{\textsc{c}2}$
\begin{eqnarray}
\bm u_{\textsc{c}1}&=&
\begin{pmatrix}
-1 & -1 & 1 & 1 & 1 & 1 
\end{pmatrix}, \\
\bm u_{\textsc{c}2}&=&
\begin{pmatrix}
-1 & -1 & 1 & 1 
\end{pmatrix}.
\end{eqnarray}
These two left null vectors finally give the two conservation laws associated with each connected component as
\begin{eqnarray}
\bm \ell^{(3)}_{\textsc{c}1}&=&
\begin{pmatrix}
1 & 1 & 1 & 1
\end{pmatrix} , \\
\bm \ell^{(3)}_{\textsc{c}2}&=&
\begin{pmatrix}
1 & 1 
\end{pmatrix}.
\end{eqnarray}
Given the direct sum structure, this can finally be summarized as 
\begin{equation}
    \bm \ell^{(3)}=
    \begin{bmatrix}
        1 & 1 & 1 & 1 & 0 & 0\\
        0 & 0 & 0 & 0 & 1 & 1
    \end{bmatrix}.
\end{equation}
We emphasize that the block diagonal decomposition of matrix $\tilde {\bm L}$ is helpful to understand how this matrix shapes the conservation law of device 3. However, for practical purposes one can rely on looking for the left null eigenvector of matrix $\bm L$ and extract the last $|\mathscr{P}^{(3)}|$ components to obtain conservation laws.
\section{Conclusion} 

The connection of thermodynamic devices, converting physical quantities of any kind, opens many possibilities that must be further explored. Our theory of thermodynamic circuits indicates promising directions for a systematic computation of conservation laws and current--force characteristics beyond the linear case. The nonequilibrium conductance matrix could simplify the numerical resolution of the boundary problem for the serial association of devices by using a recursive algorithm leading to the local potential on the connection. Experimental ways of measuring a nonequilibrium conductance matrix could be investigated as well. Theoretical works are needed to extend our work to periodic steady states beyond the stationary case or to deal with the problem of multistability frequently arising in nonlinear systems. For the former extension, the notion of nonequilibrium conductance must be generalized from stationary states to periodic steady states \cite{Goupil2016_vol94, Heimburg2017vol19, Mori2023vol14}. For the latter, because multistability usually emerges in a large volume limit, one should replace graphs with hypergraphs such as those appearing in chemical reaction networks. Current fluctuations within composite devices are also of great interest as they are tightly related to conductance matrices that describe quadratic fluctuations only. This indicates that an extension of the additivity principle of Bodineau and Derrida \cite{Bodineau2004_vol92} is within reach to determine the currents statistics of a circuit from those of its sub-devices.
Finally, inspired by electronics, impedance adaptation could be generalized to allow maximal transmission across the circuits of coupled currents.

\bibliography{BasePapierI}

\end{document}